\documentclass[english,reprint,amsmath,amssymb,aps]{revtex4-1}
\pdfoutput=1
\usepackage{graphicx}
\usepackage{amsmath}
\usepackage{hyperref}
\usepackage{tabularx}
\usepackage{hyphenat}

\begin{document}

\title{Realizing Rapid, High-Fidelity, Single-Shot\\
 Dispersive Readout of Superconducting Qubits }

\author{T.~Walter}

\thanks{These authors contributed equally to this work}

\affiliation{Department of Physics, ETH Zurich, CH-8093 Zurich, Switzerland}

\author{P.~Kurpiers}

\thanks{These authors contributed equally to this work}

\affiliation{Department of Physics, ETH Zurich, CH-8093 Zurich, Switzerland}

\author{S.~Gasparinetti}

\affiliation{Department of Physics, ETH Zurich, CH-8093 Zurich, Switzerland}

\author{P.~Magnard}

\affiliation{Department of Physics, ETH Zurich, CH-8093 Zurich, Switzerland}

\author{A.~Poto\v{c}nik}

\affiliation{Department of Physics, ETH Zurich, CH-8093 Zurich, Switzerland}

\author{Y.~Salath\'e}

\affiliation{Department of Physics, ETH Zurich, CH-8093 Zurich, Switzerland}

\author{M.~Pechal}

\affiliation{Department of Physics, ETH Zurich, CH-8093 Zurich, Switzerland}

\author{M.~Mondal}

\affiliation{Department of Physics, ETH Zurich, CH-8093 Zurich, Switzerland}

\author{M.~Oppliger}

\affiliation{Department of Physics, ETH Zurich, CH-8093 Zurich, Switzerland}

\author{C.~Eichler}

\affiliation{Department of Physics, ETH Zurich, CH-8093 Zurich, Switzerland}

\author{A.~Wallraff}

\affiliation{Department of Physics, ETH Zurich, CH-8093 Zurich, Switzerland}

\date{\today}
\begin{abstract}
The speed of quantum gates and measurements is a decisive factor for
the overall fidelity of quantum protocols when performed on physical
qubits with finite coherence time. Reducing the time required to distinguish
qubit states with high fidelity is therefore a critical goal in quantum
information science. The state-of-the-art readout of superconducting
qubits is based on the dispersive interaction with a readout resonator.
Here, we bring this technique to its current limit and demonstrate
how the careful design of system parameters leads to fast and high-fidelity
measurements without affecting qubit coherence. We achieve this result
by increasing the dispersive interaction strength, by choosing an
optimal linewidth of the readout resonator, by employing a Purcell
filter, and by utilizing phase-sensitive parametric amplification.
In our experiment, we measure $98.25\%$ readout fidelity in only
48~ns, when minimizing read-out time, and $99.2\%$ in 88~ns, when
maximizing the fidelity, limited predominantly by the qubit lifetime
of 7.6~$\mathrm{\mu s}$. The presented scheme is also expected to
be suitable for integration into a multiplexed readout architecture. 
\end{abstract}
\maketitle

\section{Introduction}

High fidelity single-shot qubit readout is quintessential for realtime
quantum feedback schemes used for example in error correction~\cite{DiVincenzo2009,Barends2014},
teleportation~\cite{Bennett1993,Steffen2013}, and state initialization~\cite{Johnson2012,Riste2012}.
It is also a key element in fundamental tests of quantum mechanics,
such as loophole-free Bell tests~\cite{CHSH1969,Ansmann2009,Hanson2015}.
Experimental progress in these areas may ultimately lead to fault-tolerant
quantum computation, for which, one of the most promising platforms
is built on superconducting circuits and qubits. The standard technique
to probe these qubits relies on the state-dependent dispersive frequency
shift imposed by the qubit on a coupled resonator~\cite{Blais2004,Wallraff2005}.
While averaging was required early on to determine the qubit state~\cite{Wallraff2005},
advances in quantum limited amplification~\cite{Caves1982,Yurke1996,Castellanos2008,Clerk2010}
allowed for the observation of quantum jumps and the discrimination
of qubit states in single shot measurements~\cite{Mallet2009,Vijay2011}.
In an attempt to further improve readout fidelities and measure multiple
qubits simultaneously, Purcell filters~\cite{Reed2010,Jeffrey2014,Bronn2015b}
and broadband parametric amplifiers were developed~\cite{Mutus2014a,Macklin2015,Roy2015c}.
With these advances, state discrimination is now predominantly limited
by qubit decay during the time of measurement~\cite{Macklin2015,Krantz2016,Liu2014}.

In order to overcome this limitation, one can reduce the measurement
time, which we achieve in this work by increasing the dispersive interaction
strength and by choosing an appropriate resonator linewidth. In addition,
we use a Purcell filter to protect the qubit from radiative decay,
and efficiently detect the microwave fields with a quantum limited
phase-sensitive amplifier. In our readout experiments we discriminate
between the qubit ground and excited state with a fidelity in excess
of 98\% in less than 50~ns. Further, we show that the measurement
time can be reduced by using a shaped ``two-step'' readout pulse
that populates the resonator faster with microwave photons than a
simple square pulse \cite{Jeffrey2014,McClure2016}.

\begin{figure}
\includegraphics{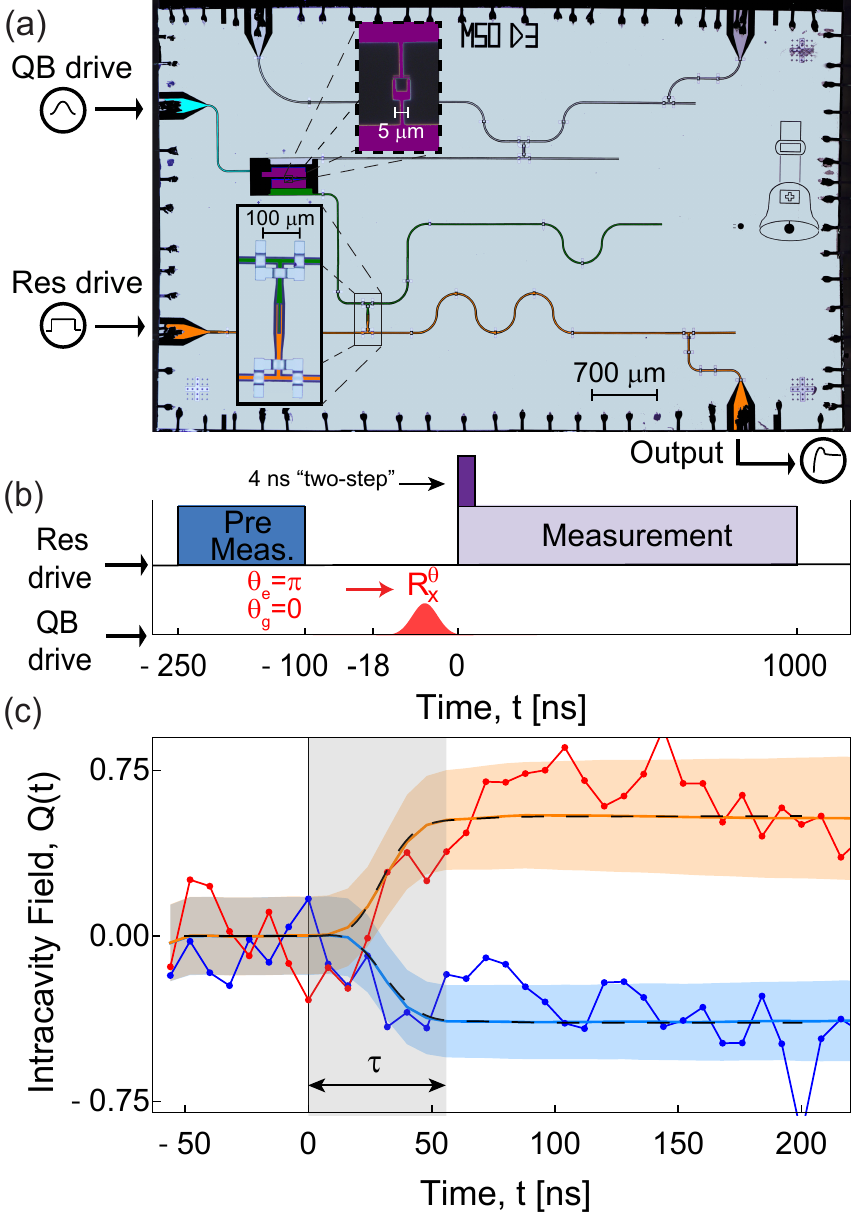} \caption{\label{fig:SetupFigures}(a) False color micrograph of the experimental
sample (see text for details). (b) Pulse scheme including the preselection
procedure, the ``gated'' (light purple) and the ``two-step'' measurement
pulse (dark+light purple), see text for details. (c) A characteristic
ground (blue) and excited (red) single-shot trajectory plotted together
with the mean of all ground and excited trajectories (light blue/
light red solid lines) with their respective standard deviations (blue
and orange shaded regions). The black dashed line shows the theoretically
expected dynamics. The gray region depicts the typical integration
time $\tau$ of the experiment.}
\end{figure}

\section{Discussion of the experiment}

Our sample device, depicted in Fig.~\ref{fig:SetupFigures}(a), consists
of a transmon qubit (purple) coupled capacitively with rate $g/2\pi=208$~MHz
to a readout resonator (green). The qubit, which is fabricated from
shadow evaporated aluminum, has a transition frequency of $\omega_{\mathrm{q}}/2\pi=$
6.316~GHz, an anharmonicity of $\alpha/2\pi=-340$~MHz and a lifetime
$T_{\mathrm{1}}=$ 7.6~$\mathrm{\mu s}$. The qubit state is controlled
by applying microwave pulses through a capacitively coupled drive
line (teal). The readout resonator, together with the remaining on-chip
elements, is fabricated from a niobium thin film sputtered on a sapphire
substrate using photolithography and reactive ion etching. The readout
resonator has a center frequency of $\omega_{\mathrm{r}}/2\pi=$ 4.754~GHz.
The detuning $\Delta=\omega_{\mathrm{q}}-\omega_{\mathrm{r}}$ is
much larger than $g$, which results in a dispersive coupling between
qubit and resonator with rate $\chi/2\pi=-7.9$~MHz. The resonator
is coupled through a Purcell filter cavity~\cite{Jeffrey2014,Sete2015}
to an external measurement line (orange structure in Fig.~\ref{fig:SetupFigures}a).
The effective resonator linewidth $\kappa_{\mathrm{eff}}=4Q_{\mathrm{p}}J^{2}/\left(\omega_{\mathrm{p}}+4\delta_{\mathrm{p}}^{2}Q_{\mathrm{p}}^{2}/\omega_{\mathrm{p}}\right)$
is controlled by the coupling strength $J/2\pi=25\,$MHz between readout
resonator and Purcell filter and the detuning $\delta_{{\rm p}}/2\pi=2\,$MHz
between the two. With the quality factor of the Purcell filter cavity
$Q_{\mathrm{p}}=74$ we find $\kappa_{{\rm eff}}/2\pi=37.5\,$MHz.

The sample is probed in a dilution refrigerator at a temperature of
10~mK using a standard microwave frequency measurement setup (Fig.~\ref{fig:Setup}a).
A Josephson Parametric Dimer (JPD) amplifier~\cite{Eichler2014a}
operated in the phase-sensitive mode is used to optimize the measurement
efficiency to achieve $\eta=0.66$, with a 3~dB bandwidth of 27~MHz
and a gain of 26~dB~\cite{Yurke1996,Clerk2010}, see Appendix~\ref{sec:JPDcalib}
for details.

To read out the qubit, we apply a coherent tone at $t=0$ to the input
port of the Purcell filter (orange in Fig.~\ref{fig:SetupFigures}(a))
at the center frequency $\omega_{\mathrm{d}}=\omega_{{\rm r}}$ between
the two state-dependent resonator frequencies. The power is chosen
to maximize the readout fidelity and corresponds to $n_{\mathrm{drive}}=2.5\pm0.25$
photons in the readout resonator, which we have calibrated based on
an ac-Stark shift measurement, see Appendix~\ref{sec:Setup-details-and}.
At $t=-250\,{\rm ns}$ we apply a $150\,{\rm ns}$ long measurement
pulse to the device, see pulse sequence in Fig.~\ref{fig:SetupFigures}(b).
We use the last $50\,{\rm ns}$ of this pulse to determine the state
of the qubit and later analyze only those traces for which the qubit
is found in the ground state. This procedure we refer to as preselection
(Appendix~\ref{sec:Preselection-And-Histogram}). Before preparing
the qubit in either the ground or excited state by applying a 18~ns
$R_{x}^{\pi}$ DRAG pulse~\cite{Motzoi2009,Gambetta2011a}, which
ends at $t=0$, we wait for about 100~ns $\approx25/\kappa_{\mathrm{eff}}$
for the resonator to decay back to its vacuum state after preselection.
In addition to the typical square (``gated'') measurement pulse
generated by gating the signal source, we also utilize a ``two-step''
pulse, which has an additional 4~ns high power segment that drives
the resonator into its steady state more rapidly~\cite{Jeffrey2014,McClure2016},
see pulse sequence in Fig.~\ref{fig:SetupFigures}(b).

A field-programmable gate array (FPGA) with an analog-to-digital converter
(ADC) samples the output signal in 8~ns time bins. The amplified
quadrature of the JPD is chosen to maximize the contrast between the
mean ground and excited state response in a single quadrature $Q(t)$.
As shown in Fig.~\ref{fig:SetupFigures}(c), the ground and excited
state response can be clearly distinguished in a single shot of a
measurement. With the qubit prepared in the excited state, due to
qubit relaxation, the averaged response $\langle Q_{e}(t)\rangle$
(solid orange line) slightly decays towards the ground state response
$\langle Q_{g}(t)\rangle$, resulting in the corresponding standard
deviation $\langle\Delta Q_{e}(t)\rangle$ (shaded orange region)
to grow in time. These trends limit the distinguishability between
ground and excited state and highlight the need for fast readout.

The theoretical dynamics of the Purcell filter~\cite{Sete2015} (dashed
black lines), including the analog and digital filtering of the measurement
line (see Section~\ref{sec:Theoretical-Model-For} and Appendix~\ref{sec:Overlap-Error-Model}),
show very good agreement with the averaged trajectories and allows
us to calibrate the starting time of the measurement. We chose $t=0$
to be the time at which the square readout pulse arrives at the input
of the Purcell filter.

In order to optimize the distinguishability between the ground and
excited state, given a certain integration time $\tau$, we evaluate
the integrated readout quadrature value $q_{\tau}=\sqrt{\kappa_{\mathrm{p}}}\int_{0}^{\tau}{\rm d}t\,Q(t)\,W(t)$,
where the weighting function $W(t)\propto\left|\langle Q_{e}(t)-Q_{g}(t)\rangle\right|$
is proportional to the average difference between the ground and excited
state responses and is normalized to $\int_{0}^{\tau}{\rm d}t\,W^{2}(t)=1$.
\begin{figure}[t]
\includegraphics{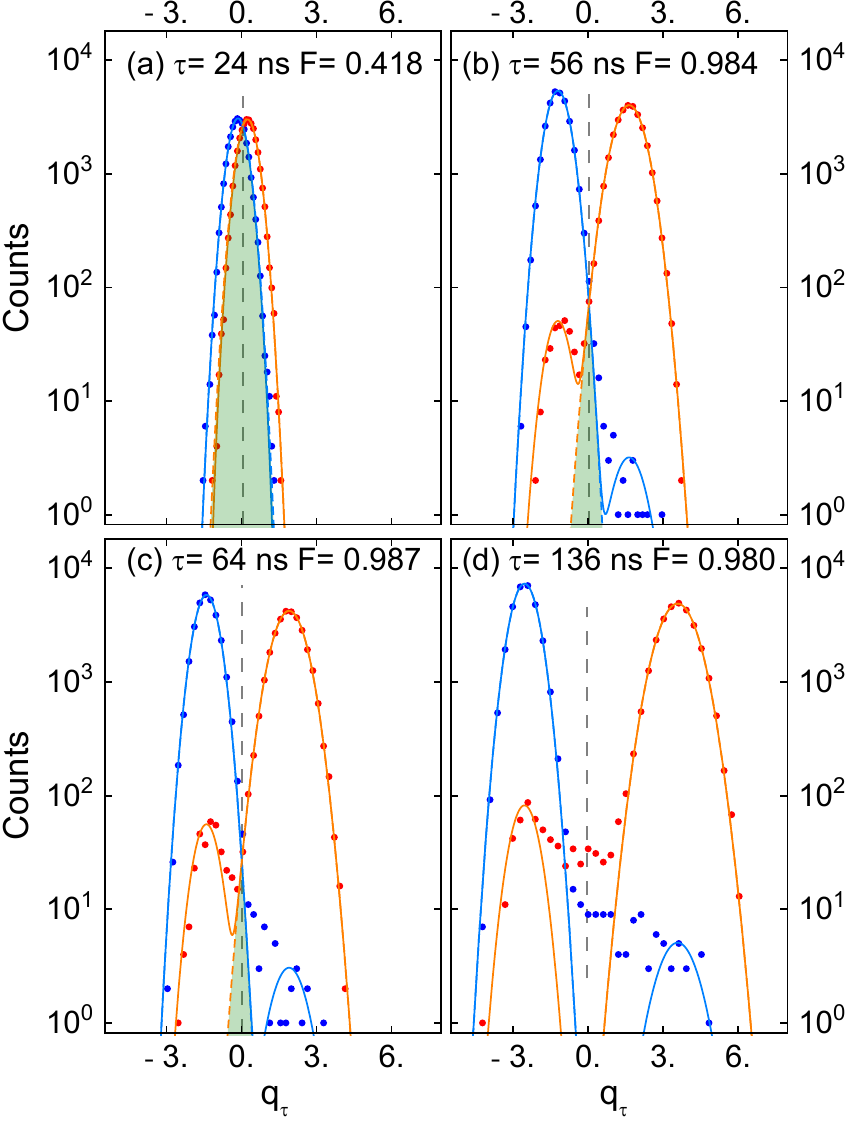}

\caption{\label{fig:SShistos}Histograms of the integrated quadrature amplitude
$q_{\tau}$ at selected integration times between 24 and 136~ns (a-d)
after preparing the qubit in the ground (blue) or excited (red) state.
The single shots were performed with a gated measurement pulse and
optimal drive power. The solid lines are fits to a double Gaussian
model whose individual components are indicated as dashed lines. The
green areas depict the overlap error. The dashed gray line indicates
the qubit state threshold. }
\end{figure}

We perform 60,000 repetitions of the experiment, alternating the qubit
preparation between ground and excited state, and populate histograms
of $q_{\tau}$, as shown in Fig.~\ref{fig:SShistos}. The ground
and excited state histograms are simultaneously fit to the sum of
two Gaussians, as described in Appendix~\ref{sec:Preselection-And-Histogram}.
From these fits we extract a decision boundary, or threshold $q_{\tau}^{\mathrm{th}}$,
at the intersection point of the two fitted state distributions. The
measurement is then characterized by its fidelity, defined as \cite{Gambetta2007}
\begin{equation}
F=1-P\left(e|g\right)-P\left(g|e\right)\label{eq:Fidelity}
\end{equation}
where $P(x|y)$ is the probability that the qubit prepared in state
$y$ is measured to be in state $x$. The fidelity is extracted directly
from the experimental data as 1 minus the sum of the fraction of ground
state preparation events with $q_{\tau}\ge q_{\tau}^{\mathrm{th}}$,
and the fraction of excited state preparation events with $q_{\tau}<q_{\tau}^{\mathrm{th}}$,
denoted $\epsilon_{\mathrm{g}}$ and $\epsilon_{\mathrm{e}}$ respectively.

At short measurement times $\tau\lesssim40\,\mathrm{ns}$, as shown
in Fig.~\ref{fig:SShistos}(a), the mean state trajectories have
yet to reach their maximum separation, leaving a large overlap of
the two distributions (green area). However, at longer integration
times the separation grows sufficiently compared to the standard deviations,
allowing us to clearly resolve the two states. With continued integration
the uncertainties further reduce, reaching a maximal distinguishability
at $\tau=64\,$ns, see Fig.~\ref{fig:SShistos}(c). For even longer
integration times, transitions between the two states during the measurement,
due to spontaneous emission, thermal excitation or transitions induced
by the readout tone, limit the fidelity.

\begin{figure}[t]
\includegraphics{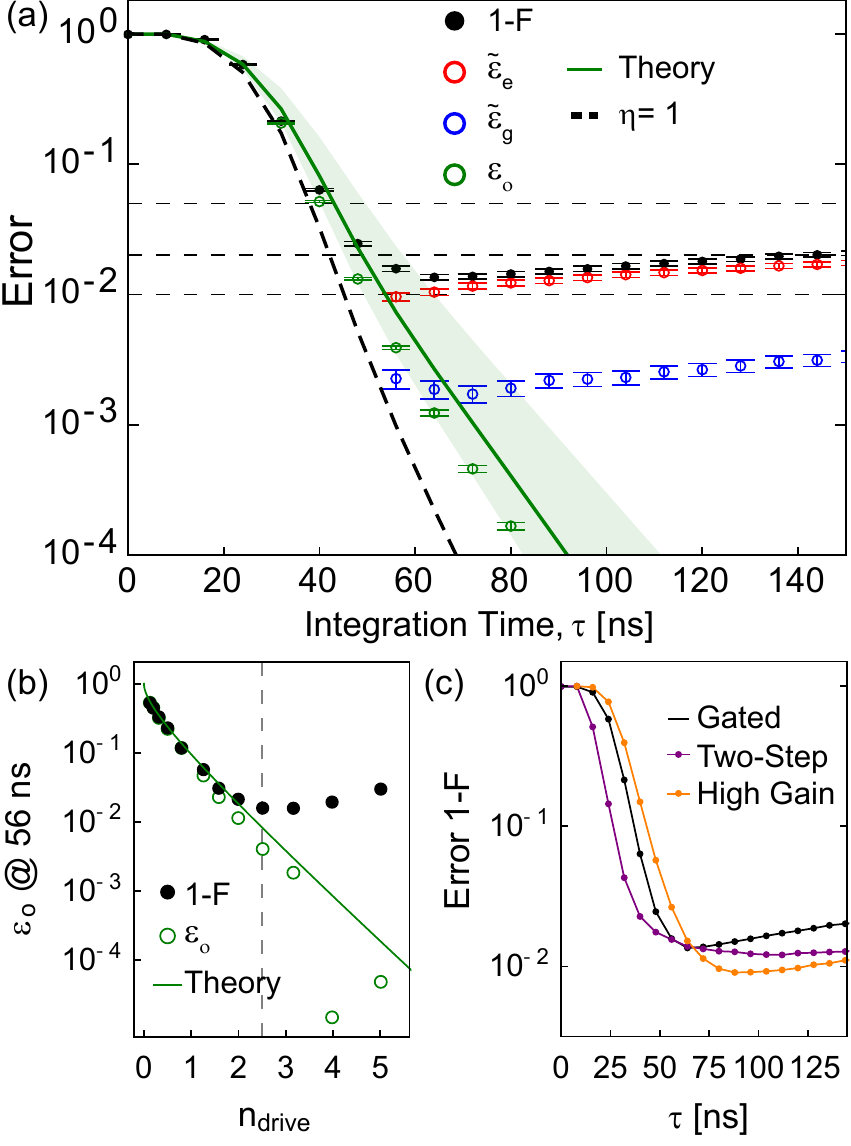}

\caption{\label{fig:ErrorsVsTime}(a) Errors $\epsilon$ of the gated single-shot
experiment at optimal drive strength $n_{\mathrm{drive}}=2.5$ as
a function of integration time $\tau$. The solid black circles represent
the infidelity extracted from the data directly. The open circles
are excited (red) ground state (blue) and overlap errors (green) extracted
from the fits. The solid green line is the numerical solution of the
full model for the overlap error (Eq.~\ref{eq:OverlapModel}) with
the independently measured system parameters including the finite
amplifier bandwidth, while the black dashed curve assumes $\eta=1$.
(b) Results of the overlap error $\epsilon_{0}$ at 56~ns extracted
from Gaussian fits (green circles) with the gated single-shot experiments
versus measurement drive power $n_{\mathrm{drive}}$. The green solid
line is the theoretical prediction using the full overlap model (Eq.~\ref{eq:OverlapModel}),
while the black dots are the measured infidelity $1-F$. (c) Experimental
results of infidelity versus integration time for the gated (black),
two-step (purple) and high amplifier gain (orange) measurements. }

\end{figure}

We further analyze the data by separating the overlap error from transition
errors. The overlap error $\epsilon_{\mathrm{o}}$ between the two
state distributions is defined as the normalized green area indicated
in Fig.~\ref{fig:SShistos} and is a measure for how distinguishable
the two distributions are \cite{Gambetta2007,Jeffrey2014}. It can
be further decomposed into contributions from the ground and excited
state $\epsilon_{\mathrm{o}}=\epsilon_{\mathrm{o,g}}+\epsilon_{\mathrm{o,e}}$.
The remaining transition errors $\tilde{\epsilon}_{x}=\epsilon_{x}-\epsilon_{\mathrm{o,x}}$,
where $x$ is the state of the qubit, are due to failed preparation
events and transitions during the measurement. The transition errors
are found to be larger for a prepared excited state and increase with
$\tau$ proportional to $1-e^{-\tau/T_{1}}$ due to spontaneous emission.
The overlap errors $\epsilon_{\mathrm{o,x}}$ decrease with integration
time, see Fig.~\ref{fig:ErrorsVsTime}. 

For $\tau=56\,\mathrm{ns}$ we obtain a fidelity of 98.42$\pm0.07$\%
and an overlap error of $\epsilon_{\mathrm{o}}=0.39\pm.01\,\mathrm{\%}$,
equal to that reported in Ref.~\cite{Jeffrey2014}, but achieved
in roughly a third of the integration time. We also have an average
assignment fidelity of 99.2\%, higher than that reported in Ref.~\cite{Bultink2016},
while using less than a fourth of the integration time. We also achieve
this higher fidelity with a qubit lifetime of 70\% and 30\% of those
in the two references~\cite{Jeffrey2014,Bultink2016}, respectively.
This further emphasizes the effectiveness of fast, optimized readout.
At this integration time, the normalized ground state error $\tilde{\epsilon}_{\mathrm{g}}=0.23\pm0.04\,\mathrm{\%}$
and normalized excited state error is $\tilde{\epsilon}_{\mathrm{e}}=0.96\pm0.06\,\mathrm{\%}$.
The limited qubit lifetime $T_{\mathrm{1}}=7.6\,\mathrm{\mu s}$ accounts
for $76\%$ of the excited state error. We attribute the remaining
part to measurement induced mixing~\cite{Gambetta2006}. This hypothesis
is supported by its equivalence to the ground state error and the
low thermal occupation of the qubit (see Appendix~\ref{sec:Preselection-And-Histogram}).
This mixing error is a consequence of driving the system stronger
in order to reduce the overlap faster and reach a higher fidelity
in a shorter time. However, some errors may be introduced by imperfect
state preparation and therefore this estimate for the measurement
induced mixing is an upper-bound.

We also performed the gated measurements at different drive strengths
and found the overlap error decreases monotonically with the measurement
power, Fig.~\ref{fig:ErrorsVsTime}(b). This behavior is in good
agreement with theory~\cite{Gambetta2007} (Appendix~\ref{sec:Overlap-Error-Model}).
However, the overall error of the readout protocol reaches a minimum
at $n_{\mathrm{drive}}=2.5$ before rising again due to measurement
induced mixing.

To further improve the readout speed at the optimal power, we repeated
the experiment with the two-step pulse shape introduced earlier~\cite{Jeffrey2014,McClure2016},
which more rapidly drives the resonator to its steady-state. The readout
fidelity during the rise time of the signal was consistently 8~ns
faster than in the gated measurements, as shown in Fig.~\ref{fig:ErrorsVsTime}(c),
and reaches a nearly identical fidelity of $F=98.25\pm.05\%$ in only
48~ns.

We also studied the dependence of readout fidelity on the measurement
efficiency $\eta$ by increasing the gain of the JPD to 35~dB. This
caused a reduction of the amplifier bandwidth to 10~MHz, but increased
the efficiency to $\eta=0.75$ (see Appendix~\ref{sec:JPDcalib}).
In this configuration we measured the highest fidelity at $99.2\%$,
however, the integration time of 88~ns required to achieve this fidelity
is longer due to the reduced bandwidth, see orange line in Fig.~\ref{fig:ErrorsVsTime}(c).

\section{Model Used for Optimizing the Readout\label{sec:Theoretical-Model-For}}

\subsection{Dynamics of readout circuit }

For phase-sensitive amplification the information acquired about the
qubit state is proportional to the difference between the ground and
excited state response in one quadrature $S(t)=\sqrt{\kappa_{\mathrm{p}}}\left|Q_{e}(t)-Q_{g}(t)\right|$
where $\kappa_{\mathrm{p}}$ is the linewidth of the Purcell filter~\cite{Sete2015}.
This is approximately equal to $S(t)\approx\sqrt{\kappa_{\mathrm{p}}}\left|\beta_{e}(t)-\beta_{g}(t)\right|$
were, $\beta_{g,e}$ are the complex intracavity field amplitudes
of the Purcell filter when the qubit is in the ground or excited state.
In the quasisteady state limit (QSS) given by $J\ll\kappa_{\mathrm{p}}$,
the full system dynamics can be reduced to the dynamics of the readout
resonator, such that $S(t)\approx\sqrt{\kappa_{\mathrm{eff}}}\left|\alpha_{e}(t)-\alpha_{g}(t)\right|$,
where $\alpha_{e,g}$ are the complex intracavity field amplitudes
of the readout resonator. By considering $\omega_{{\rm d}}=\omega_{{\rm r}}=\omega_{p}$,
for a gated measurement pulse, we find an analytical solution for
\begin{figure}
\includegraphics{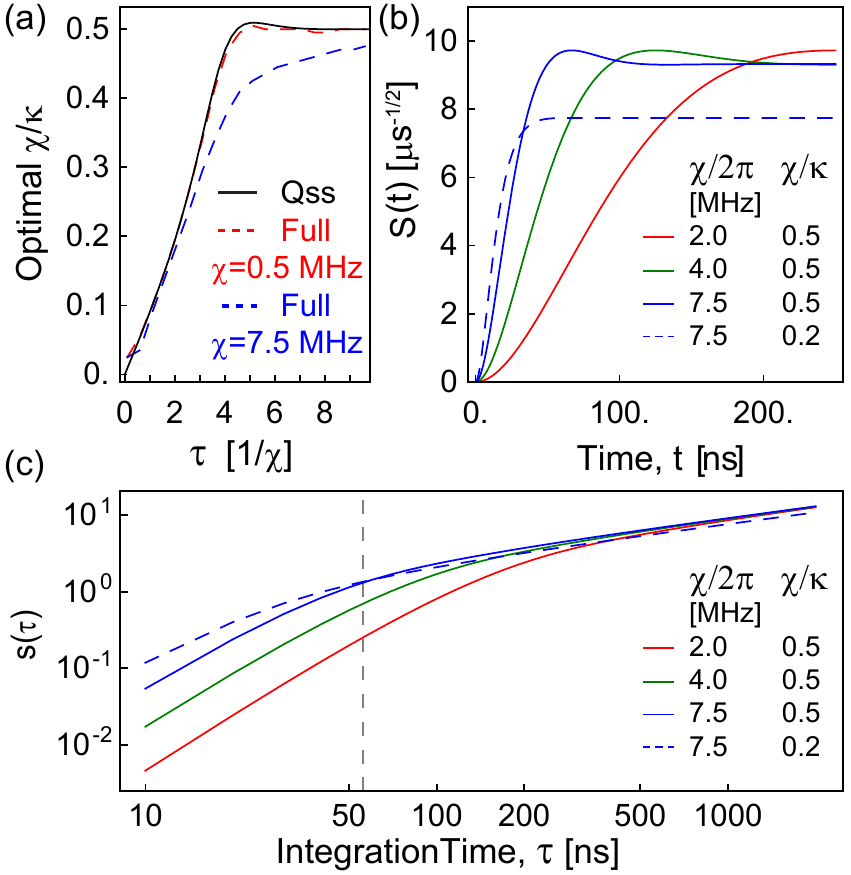}\vspace{-1.2mm}

\caption{\label{fig:SNRplot}(a) Ratio $\chi/\kappa$ that maximizes $s(\tau)$
as a function of integration time $\tau$. The solid black line represents
the quasi-steady state solution, while the dashed lines are the numerical
solutions with our circuit parameters with the full model \cite{Sete2015}
which includes the dynamics of the Purcell filter. (b) Measurement
signal $S(t)$ as a function of time $t$ for the indicated values
of $\chi$. The solid lines indicate a ratio $\chi/\kappa=0.5$, while
the dashed line is for our experimentally used ratio of 0.2, chosen
for being optimal for short integration times below 50~ns. (c) Integrated
signal $s(\tau)$ as a function of integration time $\tau$, for the
indicated values of $\chi$ and $\chi/\kappa$. The gray dashed line
represents the typical integration time used in our experiment. For
both $S(t)$ and $s(\tau)$ we have used $\alpha/2\pi=-340\,\mbox{\ensuremath{\mathrm{MHz}}}$,
$\Delta/2\pi=1562\,\mathrm{MHz}$, and $n_{\mathrm{drive}}=n_{\mathrm{crit}}/5$. }
\end{figure}

\begin{multline}
S(t)\approx\sqrt{\frac{n_{\mathrm{drive}}}{n_{\mathrm{crit}}}\left|\frac{\alpha}{\left(1+\alpha/\Delta\right)}\right|\frac{1}{|\chi/\kappa_{\mathrm{eff}}|\left(1+\left(2\chi/\kappa_{\mathrm{eff}}\right){}^{2}\right)}}\\
\left(2\left|\frac{\chi}{\kappa_{\mathrm{eff}}}\right|+e^{-\pi\kappa_{\mathrm{eff}}t}\sin(2\pi\chi t)-2e^{-\pi\kappa_{\mathrm{eff}}t}\cos(2\pi\chi t)\right)\label{eq:IntSNR}
\end{multline}
where $\chi$ is 
\begin{equation}
\chi=\frac{g^{2}}{\Delta}\frac{\alpha}{\left(\Delta+\alpha\right)}=\frac{\alpha}{4n_{\mathrm{crit}}}\frac{1}{\left(1+\frac{\alpha}{\Delta}\right)}\label{eq:chi}
\end{equation}
and $n_{\mathrm{crit}}=\Delta^{2}/4g^{2}$ is the readout resonator's
critical photon number~\cite{Blais2004}.

The readout fidelity after a measurement time $\tau$ is directly
related to the integrated information rate $s(\tau)=\left(1/\sqrt{\tau}\right)\int S(t)\,\mathrm{d}t$
(see Eq.~\ref{eq:OverlapModel}). In the following, we therefore
study $s(\tau)$ and optimize it with respect to the accessible system
parameters.

We first consider the influence of the effective linewidth $\kappa_{\mathrm{eff}}$
at otherwise constant parameters $\alpha,\,n_{\mathrm{drive}}/n_{\mathrm{crit}},\,\Delta$
and $\chi$. For integration times well into steady state we find
$s(\tau)$ to be maximized for $|\chi/\kappa_{\mathrm{eff}}|=0.5$
independent of all other parameters~\cite{Gambetta2007}, see Fig.~\ref{fig:SNRplot}(a).
However, at integration times $\chi\tau\lesssim4.5$ one could achieve
a higher fidelity by increasing $\kappa_{\mathrm{eff}}$ with respect
to $\chi$. This increases $S(t)$ at short times while reducing its
steady state value. We note that, this initial speed up could also
be achieved by using a two-step pulse while maintaining $\chi/\kappa=0.5$
independent of the integration time.

When $\chi$ and therefore $\kappa_{\mathrm{eff}}$ becomes comparable
to $\kappa_{\mathrm{p}}$, one must solve the complete system dynamics
beyond the QSS solution including the dynamics of the Purcell-filter~\cite{Sete2015}.
We have plotted the full solution with the experimentally realized
system parameters as the blue dashed line in Fig.~\ref{fig:SNRplot}(a)
which shows a slower convergence to the steady state ratio than the
QSS solution.

For the optimal ratio $\chi/\kappa_{{\rm eff}}=0.5$, we also see
from Fig.~\ref{fig:SNRplot}(a) that the time required to reach steady
state is inversely proportional to $\chi$. This emphasizes that in
order to achieve fast readout a large $\chi$ is favorable. Therefore,
if one can maintain the optimal ratio, then larger $\chi$ will more
rapidly accumulate signal. By successfully realizing these concepts
in the experiment, we were able to achieve a higher single-shot fidelity
at shorter integration times than previously reported, where it is
less affected by qubit relaxation time $T_{1}$. We chose the ratio
$|\chi/\kappa_{\mathrm{eff}}|=0.2$ in our experiment to optimize
for an integration time of 50~ns or less using the gated measurement
pulse. This is illustrated in Fig.~\ref{fig:SNRplot}(b-c) where
the integrated signal is larger for $|\chi/\kappa_{\mathrm{eff}}|=0.2$
than for $|\chi/\kappa_{\mathrm{eff}}|=0.5$ for $\tau<50\,{\rm ns}$.

\subsection{Constraints}

The optimal set of parameters are typically subject to experimental
constraints of the system. With the use of Purcell filters~\cite{Reed2010,Jeffrey2014,Bronn2015b},
$|\chi/\kappa_{\mathrm{eff}}|\approx1/2$ can be maintained for larger
values of $\chi$, without reaching a Purcell limited qubit lifetime~\cite{Koch2007}.
In this case, the limitation for readout speed derives directly from
the upper bound of $\chi$.

For larger $\chi$, a large anharmonicity is advantageous, see Eq.~\ref{eq:chi}.
However with the use of a transmon qubit, large $\alpha$ reduces
the qubit's coherence time through charge dispersion~\cite{Schreier2008}.
Charge dispersion scales exponentially with the ratio of the Josephson
energy to the charging energy, $E_{J}/E_{C}$~\cite{Koch2007} and
therefore with the qubit frequency and anharmonicity. For our sample,
we chose an anharmonicity which results in an estimated upper-bound
for the dephasing time of 150~$\mathrm{\mu s}$ with a qubit near
6.3 GHz. Our sample, however, did not realize this upper-bound due
to other dephasing mechanisms and fabrication limitations.

The second constraint on $\chi$ derives from its inverse relationship
to the critical photon number $n_{\mathrm{crit}}\propto1/\chi$ (Eq.~\ref{eq:chi}).
The dispersive approximation remains valid only in the limit $n_{\mathrm{crit}}\gg1$,
leading to an upper bound on $\chi$. Furthermore, $n_{\mathrm{crit}}$
limits the number of photons that can be used for readout without
driving unwanted qubit transitions. The probability for such a transition
is proportional to the parameter~\cite{Blais2004,Khezri2016} 
\begin{equation}
\lambda=\sin^{2}\left(\frac{1}{2}\tan^{-1}\left(\sqrt{\frac{n_{\mathrm{drive}}+1}{n_{\mathrm{crit}}}}\right)\right)
\end{equation}
implying the constraint $n_{\mathrm{drive}}\ll n_{\mathrm{crit}}$.
As a suitable trade off between these limitations, we chose $n_{\mathrm{crit}}\approx13$
and $n_{\mathrm{drive}}\approx n_{\mathrm{crit}}/5$, resulting in
$\lambda\sim1.8\%$. In our experiment, this leads to a measurement
induced error $<0.23\%$ at $\tau=56$~ns.

Additionally, the mixing rates of the qubit states during measurement
may also depend on higher order transmon states~\cite{Boissonneault2010,Sank2016},
as well as, the power spectrum of the flux noise $\Gamma_{\uparrow\downarrow}\propto\left(n_{\mathrm{drive}}/n_{\mathrm{crit}}\right)N_{f}(\pm\Delta)$,
where $N_{f}$ is the flux noise power at detuning $\mbox{\ensuremath{\pm}}\Delta$~\cite{Boissonneault2009}.
Slichter et. al. \cite{Slichter2012} measured a $N_{f}\propto1/f$
flux noise dependency and therefore we designed a $g/2\pi\approx210\,\mathrm{MHz}$
to keep the detuning $\Delta/2\pi\approx1.6$~GHz sufficiently large
for our chosen $n_{\mathrm{crit}}$. We further utilized a positive
detuning $\Delta$ to exploit the asymmetry in $\chi$ with respect
to $\Delta$ which allows for a large $\chi/2\pi>-7.5$~MHz and remain
within these constraints. 
\begin{figure*}[t]
\begin{centering}
\includegraphics{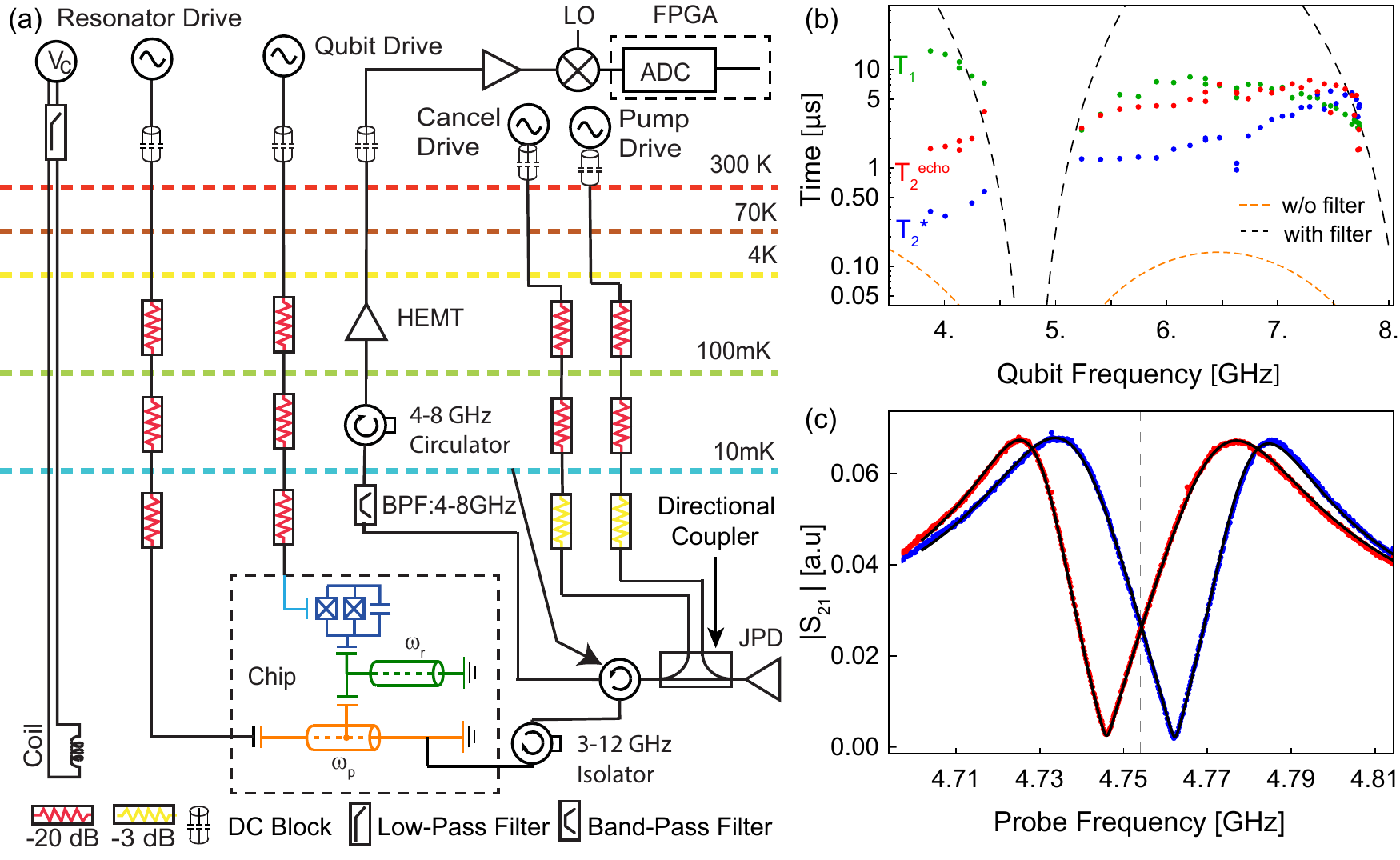} 
\par\end{centering}

\caption{\label{fig:Setup}(a) Schematic of the experimental setup. (b) Measured
lifetime $T_{1}$, Ramsey decay time $T_{2}^{\star}$, and spin echo
time $T_{2}^{\mathrm{echo}}$ as a function of the qubit frequency.
The black and orange dashed lines are the theoretical Purcell lifetimes
with and without the band pass filter. (c) Measured transmission of
readout line with the qubit in its ground (blue) and excited (red)
state. The solid black lines are fits to the input-output model of
the circuit used to extract the relevant parameters of the system. }
\end{figure*}

\section{Conclusions}

In summary, we have demonstrated an increase in speed and fidelity
of qubit readout by optimizing the circuit parameters. Further, we
have identified constraints on $\chi$ imposed by the nature of the
transmon qubit and its dispersive interaction. The limits to the qubit
lifetime and dephasing time imposed by our choice of circuit parameters
is expected to be 600~$\mathrm{\mu s}$ and 150~$\mathrm{\mu s}$
respectively, while other mechanisms not considered may impose shorter
times. We therefore believe our design concepts are extensible to
a multiplexed readout architecture~\cite{Chen2012f}.

While improvements in the qubit $T_{1}$ and in the amplifier performance
are at this point well understood to further improve the readout,
the role of measurement induced mixing as a limiting factor remains
an aspect worth further investigation.

\section*{acknowledgements}

The authors would like to thank J. Heinsoo for his efforts on improving
our automated qubit calibration software which helps characterize
our qubits and optimize the control gates. Additionally, we thank
M. Collodo, C.~K. Andersen, S. Krinner, L. Govia and A. Clerk for
discussions.

This work is supported by the European Research Council (ERC) through
the ``Superconducting Quantum Networks'' (SuperQuNet) project, by
National Centre of Competence in Research ``Quantum Science and Technology''
(NCCR QSIT), a research instrument of the Swiss National Science Foundation
(SNSF), by the Office of the Director of National Intelligence (ODNI),
Intelligence Advanced Research Projects Activity (IARPA), via the
U.S. Army Research Office grant W911NF-16-1-0071 and by ETH Zurich.
The views and conclusions contained herein are those of the authors
and should not be interpreted as necessarily representing the official
policies or endorsements, either expressed or implied, of the ODNI,
IARPA, or the U.S. Government. The U.S. Government is authorized to
reproduce and distribute reprints for Governmental purposes notwithstanding
any copyright annotation thereon.\vfill{}

\appendix

\section{Setup details \protect \\
 and sample characterization\label{sec:Setup-details-and}}

In this section we describe the measurement setup, discuss basic sample
characterization measurements and the ac-Stark shift calibration of
the readout power.
\begin{figure}
\includegraphics[width=1\columnwidth]{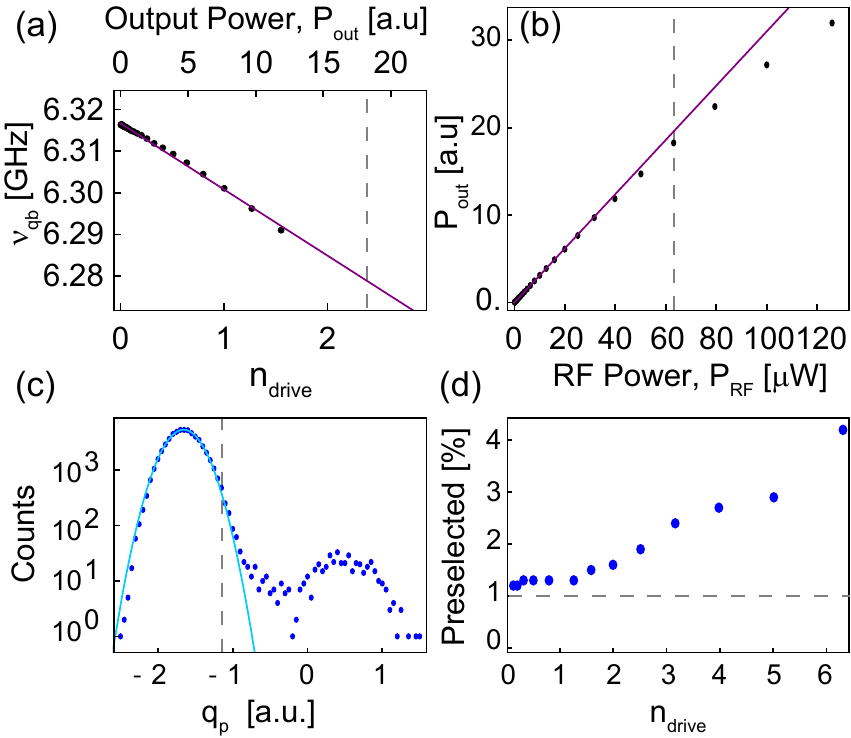} \caption{\label{fig:AcStarkandPreselect}a) Qubit frequency $\nu_{\mathrm{qb}}$
(black dots) as a function of measured power $P_{\mathrm{out}}$ with
a linear fit (purple line). The calibrated photon number is shown
on the bottom axis and the gray dashed line indicates the optimal
drive power used in the experiment. (b) Measured output power as a
function of input power $P_{{\rm RF}}$. The purple line is a linear
fit to the low power data while the gray dashed line indicates the
optimal drive power. (c) Mean value $q_{p}$ of the last 50~ns of
the preselection pulse (blue points) for $n_{{\rm drive}}=2.5$ with
Gaussian fit (blue line). The gray dashed line is the threshold used
to remove thermally excited states from further analysis set at the
99\% cumulative distribution function of the Gaussian fit. (d) Percentage
of removed data after preselection analysis as a function of drive
power. The gray dashed line indicates the minimum percent of traces
which could be removed due to our choice of the preselection state
discrimination threshold. }
\end{figure}
\begin{figure*}
\noindent \begin{centering}
\includegraphics{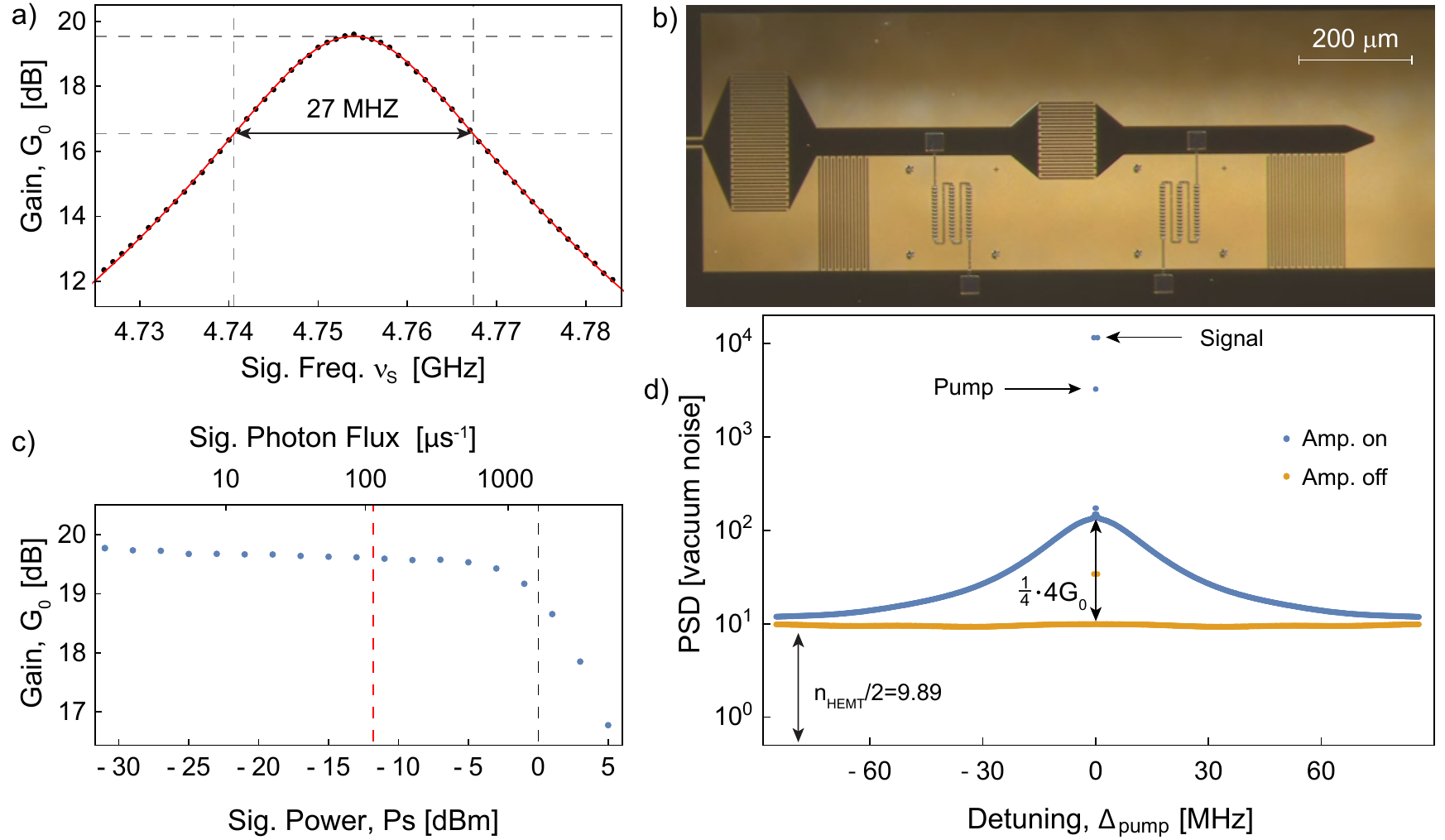} 
\par\end{centering}

\caption{\label{fig:JPDFigures}a) Measurement of the gain $G_{\mathrm{0}}$
of the amplifier versus signal frequency $\nu_{\mathrm{s}}$. The
red line is a lorentzian fit to extract the peak gain $\approx19.7\,\mathrm{dB}$
and 3~dB bandwidth of $27\,{\rm MHz}$. b) Micrograph of the Josephson
Parametric Dimer used in the experiment made from etched niobium (black)
on sapphire (yellow). The circuit consists of two capacitively coupled
lumped element non-linear resonators each with a large finger capacitor
and an array of 45 SQUIDs (shadow evaporated aluminum, gray in picture)
acting as a non-linear inductor element. The left resonator is capacitively
coupled to the output line of the measurement setup. c) Measurement
of the gain $G_{0}$ \textsl{vs.} signal power $P_{s}$ to extract
the 1dB compression point of the amplifier at the experimentally used
bias point. The red dashed line is the optimal readout power used
in the experiment while the gray dashed line depicts the 1~dB compression
power. d) Power spectral density (PSD) of the amplified quadrature
versus signal frequency detuning from the pump tone with the parametric
amplifier on (blue) and off (yellow) used to calibrate the measurement
efficiency of the output line. The y-axis is scaled such that the
distance between the peak of the parametric amplifier on spectrum
(blue) and the off noise spectrum (yellow) is equal to the the vacuum
noise times the gain, ie $1/4\cdot4G_{0}$. }
\end{figure*}

A measurement signal applied to the sample exits the output port of
the band pass Purcell filter of the sample and passes through an isolator,
a circulator and a directional coupler before entering the Josephson
Parametric Dimer (JPD), see Fig.~\ref{fig:Setup}(a). The signal
then passes through additional amplification stages before it is down
converted to 250~MHz and finally is digitized at the ADC connected
to the FPGA. The ADC samples the signal at a 1~GHz rate. Within the
FPGA the signal is digitally down-converted to DC and integrated for
8 ns in order to remove the down converted low frequency noise and
LO leakage from the analog down conversion process. The result of
the FPGA calculations are then stored for the analysis discussed in
the main text.

The implementation of a band pass Purcell filter improves the lifetime
of our qubit beyond the limit imposed by the readout resonator bandwidth
through the Purcell effect \cite{Houck2008}, shown in Fig.~\ref{fig:Setup}(b).
The measured lifetime (green dots) is between 30 and 100 times longer
than the $T_{1}$ times predicted without the Purcell filter (orange
dashed line). However, it is often shorter than the limit imposed
by the Purcell filter (black dashed line) which we attribute to limitations
in our fabrication techniques. The dephasing time of this qubit was
2$T_{1}\approx5\,\mathrm{\mu s}$ at the maximum qubit frequency ($T_{1}$
is Purcell limited here), where the sensitivity to flux noise is minimized,
and then decreases compared to $T_{1}$ when the qubit is tuned to
lower frequencies.

We characterize the sample by performing resonator spectroscopy. We
plot the transmission through the Purcell filter as a function of
probe frequency for the ground (blue) and excited (red) qubit states
in Fig.~\ref{fig:Setup}(c). From these transmission spectra, the
values of $\omega_{\mathrm{p}}$, $\omega_{\mathrm{r}}$, $J$, $\chi$,
and $Q_{p}=\omega_{p}/\kappa_{p}$ are extracted from the fit parameters
of the input-output model of the circuit given by 
\begin{equation}
\left|S_{21}\right|_{\pm}\propto\frac{\kappa_{p}}{\frac{\left(\gamma+\kappa_{p}\right)}{2}+i\left(\omega_{p}-\omega\right)+\frac{2J^{2}}{\gamma+2i\left(\omega_{r}\pm\chi-\omega\right)}}
\end{equation}
where $\gamma$ is the small internal loss rate of the Purcell filter
and readout resonator which are assumed to be equal. Furthermore,
the qubit frequency and anharmonicity are determined by standard qubit
spectroscopy~\cite{Schuster2005}.

We calibrate the photon number in the resonator $n_{{\rm drive}}$
by measuring the qubit frequency as a function of measurement drive
power $P_{\mathrm{RF}}$. Here, $n_{\mathrm{drive}}$ is assumed to
be proportional to the power $P_{\mathrm{out}}$ transmitted through
the device and measured at the FPGA, see Fig.~\ref{fig:AcStarkandPreselect}(a).
The proportionality factor between output power and intra-cavity photon
number is extracted using the known $\chi$. Due to a self-Kerr non-linearity,
the output power is proportional to the input power only at low readout
powers (Fig.~\ref{fig:AcStarkandPreselect}(b)).

\section{Preselection and histogram analysis\label{sec:Preselection-And-Histogram}}

Preselection rejects instances in which the qubit is detected in the
excited state from further analysis. Not doing so would lead to systematic
errors in the experiment. Excited state detection could be due to
thermal excitation, or residual excitation of the qubit from previous
measurement runs. In our experiment we integrate the last 50~ns of
the premeasurement pulse resulting in a mean integrated measurement
value $q_{p}$ to determine the state of the qubit. The histogram
of these integrated values, Fig.~\ref{fig:AcStarkandPreselect}(c),
is fit to a Gaussian distribution (light blue line) and the state
discrimination threshold (dashed vertical line) is set to the 99\%
of the fitted cumulative distribution function. This guarantees that
the majority of thermally excited states identified by integrated
measurement values larger than the threshold are rejected.

We find that the fraction of rejected events increases with resonator
drive strength $n_{{\rm drive}}$, see Fig.~\ref{fig:AcStarkandPreselect}(d).
This is due to measurement induced mixing which is expected to vanish
for decreasing drive strength. For $n_{{\rm drive}}\rightarrow0$
we find an excess of 0.3$\pm.1$\% of events rejected beyond the set
threshold of 1\%, which we attribute to the residual thermal population
of the qubit, see Fig.~\ref{fig:AcStarkandPreselect}(d). This low
thermal population is a byproduct of the Purcell filter, which due
to its limited bandwidth and detuning from the qubit, further reduces
the effective thermal photon flux in the readout resonator at the
qubit frequency. This is an additional benefit of using a Purcell
filter to improve the fidelity of single-shot readout not previously
discussed in Ref.~\cite{Jeffrey2014}.

For the analysis of the measurement errors as discussed in the text,
the model used to fit the single-shot histograms is given by 
\[
\left[\begin{array}{c}
C_{\mathrm{g}}\left(q_{\mathrm{\tau}}\right)\\
C_{\mathrm{e}}\left(q_{\mathrm{\tau}}\right)
\end{array}\right]=\left[\begin{array}{cc}
A_{\mathrm{gg}} & A_{\mathrm{eg}}\\
A_{\mathrm{ge}} & A_{\mathrm{ee}}
\end{array}\right]\left[\begin{array}{c}
\mathrm{PDF}\left[\mathcal{N}\left(\mu_{\mathrm{g}},\sigma_{\mathrm{g}}\right),\,q_{\mathrm{\tau}}\right]\\
\mathrm{PDF}\left[\mathcal{N}\left(\mu_{\mathrm{e}},\sigma_{\mathrm{e}}\right),\,q_{\mathrm{\tau}}\right]
\end{array}\right]
\]
where $\mathrm{PDF}\left[\mathcal{N}\left(\mu_{x},\sigma_{x}\right)\right]$
is a normal probability distribution function with mean and standard
deviation, $\mu_{x},\,\sigma_{x}$ and $C_{x}$ are the expected counts
of traces prepared in qubit state $x\in\left\{ g,\,e\right\} $ as
a function of the filtered quadrature value $q_{\mathrm{\tau}}$.
$A_{if}$ are the amplitudes with the qubit prepared in the desired
state $i\in\left\{ g,\,e\right\} $ and detected in the final state
$f\in\left\{ g,\,e\right\} $. This model does not accurately account
for transitions that occur during the measurement but only for ones
that occurred between state preparation and the beginning of the measurement.
For the short measurement times considered here this is a reasonably
good approximation.

\section{Calibration of measurement efficiency \label{sec:JPDcalib}}

We perform power spectral density measurements of the noise added
by the amplifiers in the output line to calibrate the measurement
efficiency $\eta=\eta_{\mathrm{\phi amp}}\eta_{\mathrm{loss}}$ of
our output line characterized by the parametric amplifier's noise
performance $\eta_{\mathrm{\phi amp}}$ and the loss before the amplifier
$\eta_{\mathrm{loss}}$. We achieve parametric gain by pumping the
amplifier \cite{Eichler2014a} with a coherent tone at the readout
frequency. The maximum gain is adjusted by varying the pump power
and the flux bias through the superconducting quantum interference
device (SQUID) arrays. The pump tone is interferometrically canceled
by adjusting the phase and amplitude of a cancellation tone coupled
via the directional coupler to the output of the JPD. This minimizes
pump leakage back toward the sample and avoids saturation of following
amplifiers along the output line. The frequency dependent gain peaks
at $G_{0}=$19.7~dB, see Figure~\ref{fig:JPDFigures}(a), at the
desired signal frequency of 4.754~GHz used in the experiment. By
sweeping the signal power we determine the 1~dB compression point
of the amplifier (Fig.~\ref{fig:JPDFigures}(c)), which is about
10~dB higher than the power used for measuring the qubit state in
our experiment.

Next, we characterize the amplifier's measurement efficiency when
operating in the phase-sensitive mode $\eta_{\mathrm{\phi amp}}$,
i.e amplifying and detecting only one quadrature of the signal field.
In order to create a single quadrature signal distinguishable from
the pump we apply two coherent signals of equal power and opposite
detuning from the pump tone. We then adjust the phase of the two tones
so that the resulting signal is oriented along the amplified quadrature,
achieving a phase sensitive gain, $4G_{0}$. For this measurement,
the ADC samples the output signal for 8.192~$\mu$s with 1~ns resolution,
resulting in a frequency resolution of approximately 112~kHz. We
therefore chose to detune the signals 488~kHz above and below the
pump, which is an integer multiple of this resolution. Finally, we
measure the noise performance of the amplified quadrature by measuring
its power spectral density (PSD). The down conversion local oscillator
phase is also adjusted to maximize the gain along the real quadrature
of the complex signal entering the ADC as is done for the main experiment.
We use the FPGA to take the real component of the complex signal and
compute the auto-correlation function and average the results over
a million times. The Fourier transform of this data then reveals the
noise spectral density of this quadrature as seen in Figure~\ref{fig:JPDFigures}(d).
When the JPD is on, the overall noise level increases compared to
when it is off following the frequency dependent gain of the amplifier.
The two coherent signal peaks can be easily distinguished from the
noise floor in both the amplifier on and off spectra, as well as the
pump tone for the amplifier on case.

With the assumption that the noise of the parametric amplifier is
dominated by amplified vacuum noise, the spectral density measurements
are scaled so that the noise increase when turning on the JPD is equal
to $4G_{0}\cdot1/4$. In this expression, 1/4 represents the noise
at the input of the parametric amplifier in the amplified quadrature
and $4G_{0}$ is the gain of this quadrature. The resulting noise
offset, $n_{\mathrm{HEMT}}/2$ is 9.89 times larger than the vacuum
level. This offset is predominately due to the noise added by the
following HEMT amplifier at the 4~K temperature stage and is divided
by 2 to account for the single quadrature measurement. Following Frii's
formula for noise performance of cascaded amplifier chains \cite{Pozar2012},
this added noise will be divided by the gain of the parametric amplifier
resulting in $\eta_{\mathrm{\phi amp}}=\left(1+n_{\mathrm{HEMT}}/(2G_{0})\right)^{-1}$.
We estimate that with a gain of $4G_{0}=$ 26~dB, the noise performance
is about $\eta_{\mathrm{\phi amp}}=0.92$. If we adjust the pump settings
to increase the gain to $4G_{0}=35\,\mathrm{dB}$, at the cost of
bandwidth, we find the amplifier can achieve $\eta_{\mathrm{\phi amp}}=0.99$.

Finally, we determine the transmission efficiency $\eta_{\mathrm{loss}}$
between the output of the sample and the input of the parametric amplifier
by comparing the coherent signal power in the scaled PSD measurement
to the expected power calibrated with the AC stark shift measurement,
i.e. $P_{{\rm out}}=\left(\chi/J\right)^{2}\kappa_{{\rm p}}n_{{\rm drive}}$.
The $\left(\chi/J\right)^{2}$ prefactor is a result of driving the
circuit through the input of the Purcell filter and not the readout
resonator directly~\cite{Sete2015}. The ratio between the scaled
PSD measurement signal power and $P_{{\rm out}}$ is equal to the
transmission efficiency and found to be $\eta_{\mathrm{loss}}\approx0.75$
corresponding to approximately 1.3 dB attenuation. This is well within
the specified insertion loss of the elements between the sample and
amplifier. The measured total efficiency of the line is determined
to be $\eta=\eta_{\mathrm{loss}}\eta_{\mathrm{\phi amp}}=0.66$ for
26~dB gain and $\eta=0.75$ for 35~dB gain.

\section{Overlap error model\label{sec:Overlap-Error-Model} }

We model the readout as a measurement of a random variable from the
qubit state Gaussian distributions which have opposite means $\pm S(t)/2$
(see Sec\ref{sec:Theoretical-Model-For}) and equal variance, $\sigma^{2}(t)=1/(4\eta)$
at time $t$. With no state transitions, the optimal threshold is
at the intersection of the two distributions and including both analog
and digital filtering we find 
\begin{equation}
\epsilon_{\mathrm{o}}(\tau)=\mathrm{erfc}\left[\sqrt{\frac{1}{8}}\frac{\int_{0}^{\tau}S(t)\,f(\tau-t)\,dt}{\sqrt{\int_{0}^{\tau}\sigma^{2}(t)\,f(\tau-t)\,dt}}\right]\label{eq:OverlapModel}
\end{equation}
where $f(\tau-t)$ is the response function of the relevant filter
elements for the measurement of $s_{\tau}$. In our experiment, $f(\tau-t)$
consists of the narrow bandwidth parametric amplifier, the 8~ns box
car integration of the FPGA and the mode-matched digital weighting
function $W(t)$. We use this model along with the numerical solutions
of $S(t)$ using the full system equations (no quasi-steady state
assumption~\cite{Sete2015}), to determine the theory curves (green)
in Fig.~\ref{fig:ErrorsVsTime}(a-b) of the main text.

\bibliographystyle{apsrev4-1}
\bibliography{RefDB/QudevRefDB}

\end{document}